\documentclass[11pt,a4paper]{article}
\usepackage{times}
\usepackage[top=3.5cm,
				bottom=1.63cm,
				left=2.0cm,
				right=2.0cm,
				includefoot]{geometry}
\usepackage{psfrag}
\usepackage{amsmath,amsfonts,amsthm}

\title{Fusion Discrete Logarithm Problems}

\author{
\small Martin Schaffer \\
\small Business Unit Identification\\
\small NXP Semiconductors Austria GmbH\\
\small  \texttt{martin.schaffer@nxp.com} \and \small Stefan Rass\\
\small System Security Group \\ 
\small Klagenfurt University, Austria \\ 
\small \texttt{s.rass@syssec.at}}
\date{February 2010}
\begin{document}

\maketitle 

\newcommand{\f}[1]{\textsf{#1}}
\newcommand{\EF}{\lambda}
\newcommand{\G}{\mathbb{G}}
\newcommand{\F}{\mathbb{F}}
\newcommand{\Z}{\mathbb{Z}}
\newcommand{\N}{\mathbb{N}}
\newcommand{\MOD}{\;\textrm{MOD}\;}
\theoremstyle{plain}
\newtheorem{thm}{Theorem}
\newtheorem{exl}[thm]{Example}
\newtheorem{cor}{Corollary}
\theoremstyle{definition}
\newtheorem{defn}{Definition}
\theoremstyle{remark}
\newtheorem{rem}{Remark}

\newcommand{\TODO}[1]{\textcolor{red}{\textbf{TODO:} #1}}
\begin{abstract}
The Discrete Logarithm Problem is well-known among cryptographers,
for its computational hardness that grants security to some of
the most commonly used cryptosystems these days. Still, many of
these are limited to a small number of candidate algebraic structures
which permit implementing the algorithms. In order to extend the applicability
of discrete-logarithm-based cryptosystems to a much richer
class of algebraic structures, we present a generalized form of exponential
function. Our extension relaxes some assumptions on the
exponent, which is no longer required to be an integer. Using an axiomatic
characterization of the exponential function, we show how to
construct mappings that obey the same rules as exponentials, but can
raise vectors to the power of other vectors in an algebraically sound
manner. At the same time, computational hardness is not affected (in fact, 
the problem could possibly be strengthened).
Setting up standard cryptosystems in terms of our generalized exponential function 
is simple and requires no change to the existing security proofs. 
This opens the field for building much more general schemes than the
ones known so far.
\end{abstract}
\section{Introduction}\label{sec:intro}

Many cryptosystems are proven to be secure under a particular
computational assumption, such as RSA \cite{rsa78} for instance,
resting its security on the difficulty of the Factoring Problem.
 Many others, such as ElGamal \cite{e85,e85b}, are based on the Discrete Logarithm Problem \cite{m90} and
 other related problems on which the focus of this paper lies. Henceforth, we consider
 a group $\G_q$ of prime order $q$, for simplicity. Therein, the \emph{Discrete Logarithm Problem} (DLP) is the following: given $y,g\in
 \G_q$, $g\neq 1$, and $q$, find $x\in \Z_q$, such that $y=g^x$. The integer $x$ is called the \emph{discrete logarithm} of $y$
 to the base $g$, here denoted as dlog$_g(y)$. The problem of constructing $g^{x_1x_2}$ solely from $(g^{x_1},g^{x_2})$ is
 known as the \emph{Diffie-Hellman Problem} (DHP) \cite{dh76}. To decide, whether a given triple $(y_1,y_2,y_3)\in
 \G_q^3$ is of the form $(g^{x_1},g^{x_2},g^{x_1x_2})$ is known as the \emph{Decision Diffie-Hellman
 Problem} (DDP) \cite{b98}. Obviously, solving the DLP gives trivial solutions to the DHP and the DDP, respectively.
 Similarly, solving the DHP leads to an efficient solution of the DDP. The inverse directions are less obvious.
The DLP and the DHP have been shown to be computationally equivalent for groups whose order is free
of multiple large prime factors \cite{mw98, mw99}. This is the case for $\G_q$, since $q$ is prime.
In contrast to that, the computationally equivalence between the DHP and the DDP has only shown
to hold for groups whose order only consists of small primes factors \cite{mw98b}. In contrast, the  computationally equivalence between
the DHP and DDP for $\G_q$, $q$ prime, has not been shown yet.

Several cryptosystems are based on Discrete Logarithm Problems. The ElGamal encryption scheme, for instance, is
 semantically secure under the assumption that solving the DHP is
 hard. Moreover, under the assumption that the DDP is hard, it is guaranteed that upon two given ciphertexts, it is not efficiently
 possible to decide, if both contain the same plaintext. Unfortunately, the ElGamal
 encryption scheme is insecure against chosen
 ciphertext attacks \cite{s98}. The Cramer-Shoup
 encryption scheme \cite{cs98,cs04} overcomes this drawback, while resting its security on the DDP. 
Besides encryption schemes, the DLP can be found in several signature schemes, such as in ElGamal's \cite{e85} 
or Schnorr's \cite{s89}, or in the Digital Signature Algorithm \cite{fipspub186}. Interactive proofs of knowledge \cite{bg93} (in
particular $\Sigma$-proofs \cite{c96}), commitment schemes \cite{bcc88,cd97,dam99}, verifiable encryption \cite{s96}, verifiable secret sharing \cite{fm87,p91vss,gjkr07}, and secure multi-party computation \cite{grr98} belong to advanced
cryptographic techniques, that are often based on Discrete Logarithm Problems. Especially the
DDP finds wide attraction in applications where privacy plays an important role, such as in voting schemes \cite{cs97} or anonymous credential systems \cite{cl01}.

Recently, we managed to generalize the standard exponential function on a group $\G_q$ to a pendant that takes pairs in the base and the exponent, rather than scalars. This function shares the basic properties of exponential functions, and allows us to call it ``exponentiation''. Because all four input-elements are
 uniformly included (we call this property ``fusion'') for the computation of the output, we call this kind of exponential function ``Fusion Exponential Function'' (see end of Section \ref{sec:genfusexp:cons} for a discussion of the fusion-property). The latter also avoids confusion with ordinary exponentiation. The Fusion
 DLP (FDLP), the Fusion DHP (FDHP) and the Fusion DDP (FDDP) are defined in the usual way.
Our prelimiary results can be found
in \cite{sr08}. 

In this paper we sketch the results of \cite{sr08} in a more constructive way
and \emph{generalize} the Fusion Exponential Function such that it works with $n$-tuples of elements of $\G_q$
in the basis and $n$-tuples of elements of $\Z_q$ in the exponent, for $n$ not being restricted to $n=2$
as it is the case in \cite{sr08}. We also show that the basic properties 
are still common with ordinary exponentiation and that the latter
is a special case of the Generalized Fusion Exponential Function, i.e.~it also holds for $n=1$. Afterwards, 
we define the FDLP, FDHP and FDDP in the generalized setting and show security relations between the Discrete Logarithm Problems in the ordinary and the fusion setting. Finally, applications and possible security benefits are discussed.
\section{Exponentiation in a Group of prime Order}\label{sec:ordexp}

\subsection{Basic Properties}

As is well known, $g^x$ is defined as the $x$-fold product of $g$ with itself. For all $g,h\in \G_q$ and
$x,y\in \Z_q$ we have the following properties:
\begin{eqnarray}
(g^x)^y&=&g^{xy} \label{eq:prop1}\\
g^{x+y}&=&g^xg^y \label{eq:prop2}\\
(gh)^x&=&g^xh^x\label{eq:prop3}
\end{eqnarray}
Furthermore, $g^0=1$ and $g^{-x}=(g^x)^{-1}$. The properties stated
above are fundamental for realizing discrete-logarithm-based
cryptosystems.

\begin{rem}\label{rem:redundant}
Property \eqref{eq:prop3} is redundant, as being consequence of \eqref{eq:prop1} and \eqref{eq:prop2} and the fact that $h$ can be written as $g^w$, for $w\in\Z_q$ and $g\in\G_q\setminus\{1\}$, i.e.~
\begin{align*}
(gh)^x=(gg^w)^x\stackrel{\eqref{eq:prop2}}{=}(g^{1+w})^x\stackrel{\eqref{eq:prop1}}{=}g^{(1+w)x}=g^{x+wx}\stackrel{\eqref{eq:prop2}}{=}g^xg^{wx}\stackrel{\eqref{eq:prop1}}{=}g^x(g^w)^x=g^xh^x.
\end{align*}
 \end{rem}

\subsection{Computing Discrete Logarithms}

Computing $y=g^x$, for a given $g\in \G_q$ and $x\in \Z_q$, can be done 
efficiently. For instance, the Square-and-Multiply \cite{hoc} algorithm requires only $O(\log q)$ group operations.
However, no efficient \emph{generic} algorithm for solving the DLP is known, except for some special cases where parameters are chosen in a particular manner. A generic algorithm does not exploit any specific properties of the objects to which it is applied \cite{s97}. It works on any group, where each element can be encoded as a binary string and group operations can be considered as a black-box. One of the best known
generic attack algorithms are Pollard's rho algorithm \cite{p78} and Shank's Baby-Step-Giant-Step algorithm \cite{s73}, having an
exponential running time, lying in $O(\sqrt{q})$ and $O(\sqrt{q}\log q)$, respectively. Due to their computational complexity they are also called generic square-root attacks \cite{t01}. If the order of the group is a composite $n$,
then the best attack known to date is the Pohlig-Hellman algorithm \cite{ph78}, computing $x$ in $O(\sqrt{r})$ steps, where
$r$ is the largest prime factor of $n$. 

For a chosen group, an algorithm may exist that takes advantage of some special properties of the group. Such an algorithm is not generic since it is not applicable to any group structure. For instance, if $\G_q$ is a subgroup of $\Z_p^*$, where $p$ is a prime, the Index-Calculus algorithm \cite{hoc} can compute $x$ in sub-exponential computing time, being more efficient than a generic square-root-attack. However, this algorithm cannot be applied to $\G_q$ being a subgroup of an elliptic curve group over a Finite Field, for instance. So far, no algorithm is known that computes elliptic curve discrete logarithms faster than in $O(\sqrt{q})$ steps.
\section{Basic Fusion Exponential Function}\label{sec:fusexp}

In the \emph{fusion-setting}, as introduced in \cite{sr08}, exponents are defined as pairs of integers in
$\Z_q$. It is convenient to have the exponents of the
extended exponentiation coming from a field (in fact a commutative
ring with 1 would suffice, but a field gives rise to a wider class
of applications), while in the basis a group is most likely sufficient. A natural choice for the source of the exponents is thus a
field of order $p=q^2$, which is easily constructed by choosing
$q\equiv 3\;(\textrm{mod}\;4)$, and setting $\F_p:=\Z_q[X]\slash(X^2+1)$, for instance.
\begin{rem}
For simplicity, we sometimes denote a pair $(a,b)\in\Z_q^2$ or $(a,b)\in\G_q^2$ by a sans-serif letter, say $\f{x}$, for instance. 
\end{rem}
Let us review the derivation of the basic Fusion Exponential Function, as given in \cite{sr08}\footnote{There, referred to as \emph{Fusion-Exponentiation}.}. This idea will later be amended to yield the general scheme. To realize schemes based on the Diffie-Hellman paradigm \cite{dh76}, any exponential function candidate needs to obey property \eqref{eq:prop1} at least, so let us define a simple form of generalized exponential function, taking a pair in the exponent as
\begin{equation}\label{eq:fexp-convention}
g^{\textsf{x}} = g^{(c,d)}:=(g^c,g^d),
\end{equation}
where $\textsf{x}\in \F_p$, $\textsf{x}=(c,d)$ and $g\in
\G_q\setminus\{1\}$, thus $g$ having order $q$. Suppose we are given a term $g^{\textsf{x}}$
according to the convention \eqref{eq:fexp-convention}, and we wish
to find $(g^{\textsf{x}})^{\textsf{y}}$ such that the result equals
$g^{\textsf{x}\textsf{y}}$, i.e.~we need to calculate the latter
term given only $g^{\textsf{x}}=(g^c,g^d)$ and $\textsf{y}=(e,f)$,
where $\textsf{y}\in \F_p$. This is easily done by doing the multiplication in the exponent within $\F_p$, as
\begin{eqnarray}
    g^{\textsf{x}\textsf{y}}&=& g^{(c,d)(e,f)}\nonumber\\
   &=&g^{(ce-df,cf+de)}\nonumber\\
    &\stackrel{\eqref{eq:fexp-convention}}{=}& (g^{ce-df},g^{cf+de})\nonumber\\
&\stackrel{\eqref{eq:prop2}}{=}&\left(g^{ce}g^{-df},g^{cf}g^{de}\right)\nonumber\\
&\stackrel{\eqref{eq:prop1}}{=}&\left((g^c)^e(g^d)^{-f},(g^c)^f(g^d)^e\right)\label{eq:fexp-raw-form}.
\end{eqnarray}

Hence, we can define
\begin{equation*}
(g^{\textsf{x}})^{\textsf{y}}=(g^{(c,d)})^{(e,f)}\stackrel{\eqref{eq:fexp-convention}}{=}(g^c,g^d)^{(e,f)}
\end{equation*}
through \eqref{eq:fexp-raw-form} as
\begin{equation*}
(g^{\textsf{x}})^{\textsf{y}}
:=((g^c)^e(g^d)^{-f},(g^c)^f(g^d)^e)=g^{\textsf{x}\textsf{y}}.
\end{equation*}

Since $g$ is primitive, we can write any two elements
$a,b\in \G_q$ as $a=g^c$, $b=g^d$ for some integers $c,d\in \Z_q$.
Substituting the powers of $g$ in \eqref{eq:fexp-raw-form} gives
\begin{equation}\label{eq:fexp-substituted}
(g^{\textsf{x}})^{\textsf{y}} =(a^e b^{-f},a^fb^e),
\end{equation}
and the \emph{Fusion Exponential Function} is found by observing that by
\eqref{eq:fexp-convention}, any pair $(a,b)\in\G_q\times\G_q=:\G_p$
can be written using powers of $g$ as $(g^c,g^d)$, such that with
$g^{\textsf{x}}$ being represented by $(a,b)$, from
\eqref{eq:fexp-substituted} we arrive at the definition
\begin{equation*}
    (a,b)^{(e,f)} := (a^e b^{-f},a^fb^e),
\end{equation*}
satisfying \eqref{eq:prop1} by construction. Since $\G_p$ is simply
the direct product $\G_q^2$, it is a group with
component-wise multiplication. Having this
together with $\F_p$ being a field, the properties \eqref{eq:prop2}
and \eqref{eq:prop3} can be verfied instantly \cite{sr08}. To keep
computing discrete logarithms hard, it is intrinsic that
exponentiation is done using a basis of large order. In $\G_q$,
every element $g\neq 1$ has maximum order $q$. An analogous result
can be shown for $\G_p$ regarding the Fusion Exponential Function: every
element $\textsf{g}\neq 1$ can be used to generate $\G_p$ using
the Fusion Exponential Function, hence the corresponding (fusion) discrete logarithm as the inverse function is well-defined. A proof for the fusion-setting where $\F_p=\Z_q[X]/(X^2+1)$ can be found in \cite{sr08}.

The focus of the remainder of this paper lies in extending the above
constructive approach from $n=2$ to \emph{any} $n\geq 1$. Thus, achieving 
a definition for the \emph{Generalized Fusion Exponential Function}.
\section{Generalized Fusion Exponential Function}\label{sec:genfusexp}

In this section, we generalize the approach of Section \ref{sec:fusexp}, such that
exponents are $n$-tuples of integers in $\Z_q$ and bases are $n$-tuples of elements in $\G_q$.
\begin{rem}
Again, we sometimes denote an $n$-tuple $(x_0,\ldots,x_{n-1})\in\Z_q^n$ resp.~$(g_0,\ldots,g_{n-1})\in\G_q^n$ by the sans-serif letter $\f{x}$ resp.~$\f{g}$. In contrast to Section \ref{sec:fusexp}, the components of an $\f{x}$ are always referred to by the same letter $x_i$ using the standard font and the associated index $i$.
\end{rem}
\subsection{Vectors in the Exponent}\label{sec:vec_exp}

Let us replace the source of exponents by the Finite Field $\F_p:=\Z_q[X]/(f)$ where $f$ is an irreducible polynomial of degree $n$,
for some integer $n\geq 1$, thus having $p=q^n$. In order to provide a compact generalization of the Fusion Exponential Function, we need to consider the multiplication in $\F_p$ in more detail. Let $\f{x},\f{y}\in \F_p$, written as
\begin{align*}
\f{x}=\sum_{i=0}^{n-1} x_iX^i\quad \mathrm{and}\quad \f{y}=\sum_{i=0}^{n-1} y_iX^i
\end{align*}
for some coefficients $x_i,y_i\in \Z_q$. Without loss of generality, assume $f$ to be monic, and write
\begin{align*}
f=\left(\sum_{i=0}^{n-1} f_iX^i\right)+X^n
\end{align*}
for $f_i\in \Z_q$, $i=0,1,\ldots,n-1$, and thus obviously,
\begin{align}
X^n = \left(-\sum_{i=0}^{n-1}f_iX^i\right)\MOD{f}\label{eq:XN}.
\end{align}
Furthermore, the (plain) product $\f{z}=\f{x}\f{y}$ is of degree at most $2n$, and the $i$-th coefficient $z_i$ is given by the Cauchy-sum
\[
    z_i = \sum_{\tiny{\shortstack{$j,k\geq 0$\\$j+k=i$}}}x_jy_k
\]
for $i =0,1,\ldots,2n$. Thereby, $x_j=0$ for $j>n-1$ and $y_k=0$ for $k>n-1$. To find the remainder of $\mathsf{z} = \mathsf{xy}
= \sum_{i=0}^{2n}z_iX^i$, we can exploit the representation of $X^n$  through coefficients of
$f$ as given in \eqref{eq:XN} \cite{hk08}. This extends to higher orders by taking
\[
X^{n+1}=XX^n = X\left(-\sum_{i=0}^{n-1} f_iX^i\right)=-\sum_{i=0}^{n-1} f_iX^{i+1},
\]
which can again be decomposed
recursively to reach a representation solely via the base monomials $1$, $X$, $X^2$, $\ldots$, $X^{n-1}$. Notice, that in this
decomposition, only products of coefficients of $f$ occur, which means that by rewriting the $2n$-order polynomial
$\mathsf{z}$ in terms of $1,X,X^2,\ldots,X^{n-1}$, the resulting expressions for the coefficients become nonlinear in each
$f_i$, but remain \emph{linear} in each $x_i$ and in each $y_i$ for all $i=0,1,\ldots,n-1$ (cf.~the Cauchy sum).
Rearranging terms by pulling $x_j$, for $j=0,\ldots,n-1$, out of all products for the $i$-th coefficient of $\mathsf{xy}$ and denoting the
factor associated with $x_j$ as $\EF_{i,j}(\mathsf{y})$ (omitting the coefficient vector of $f$ because it is static) we can represent the (modulo-reduced) product $\f{xy}\text{ MOD }f$ with coefficients $z'_i$ as
\begin{equation}
    z'_i = \sum_{j=0}^{n-1}x_j\EF_{i,j}(\mathsf{y})\label{eq:prodcoeff}
\end{equation}
for $i=0,1,\ldots,n-1$. Notice, that for any fixed $f$, $\EF_{i,j}:\F_p\rightarrow \Z_q$ is a known fixed function for $j=0,1,\ldots,n-1$, where the linearity in
each coefficient of the input is inherited, thus having
\begin{equation}
    \lambda_{i,j}(\mathsf{x})+\lambda_{i,j}(\mathsf{y}) = \lambda_{i,j}(\mathsf{x}+\mathsf{y})\label{eq:lem_tilde}
\end{equation}
for all $\mathsf{x},\mathsf{y}\in\F_p$.

\begin{rem}
For simplicity, we henceforth represent polynomials through the vector over their coefficients exclusively, i.e.~we write either $(x_0,\ldots,x_{n-1})$ or $\f{x}$ instead of $\sum_{i=0}^{n-1} x_iX^i$. Addition is as usual component-wise and for multiplication we use our adapted representation
\begin{align}
(x_0,\ldots,x_{n-1})(y_0,\ldots,y_{n-1})\stackrel{\eqref{eq:prodcoeff}}{=}\left(\sum_{j=0}^{n-1}x_j\EF_{0,j}(\f{y}),\ldots,\sum_{j=0}^{n-1}x_j\EF_{n-1,j}(\f{y})\right).\label{eq:polymul}
\end{align}
\end{rem}

\subsection{Construction}\label{sec:genfusexp:cons}

Analogously to \eqref{eq:fexp-convention} we define
\begin{equation}\label{eq:genfexp-convention}
g^{\f{x}} = g^{(x_0,\ldots,x_{n-1})}:=(g^{x_0},\ldots, g^{x_{n-1}}),
\end{equation}
where $\f{x}\in \F_p$, $\f{x}=(x_0,\ldots,x_{n-1})$ and $g\in \G_q\setminus \{1\}$. In this setting we
wish to calculate $g^{\f{x}\f{y}}$, for $\f{y}\in \F_p$, given only $g^{\f{x}}=(g^{x_0},\ldots, g^{x_{n-1}})$ and $\f{y}=(y_0,\ldots, y_{n-1})$. Carrying out the multiplication $\mathsf{xy}$ in $\F_p$, we find
\begin{align}
g^{\f{x}\f{y}}=g^{(x_0,\ldots,x_{n-1})(y_0,\ldots,y_{n-1})}&\stackrel{\eqref{eq:polymul}}{=}g^{\left(\sum_{j=0}^{n-1}x_j\EF_{0,j}(\f{y}),\ldots,\sum_{j=0}^{n-1}x_j\EF_{n-1,j}(\f{y})\right)}\nonumber\\
&\stackrel{\eqref{eq:genfexp-convention}}{=}\left(g^{\sum_{j=0}^{n-1}x_j\EF_{0,j}(\f{y})},\ldots,g^{\sum_{j=0}^{n-1}x_j\EF_{n-1,j}(\f{y})}\right)\nonumber\\
&\stackrel{\eqref{eq:prop2}}{=}\left(\prod_{j=0}^{n-1} g^{x_j\EF_{0,j}(\f{y})},\ldots,\prod_{j=0}^{n-1} g^{x_j\EF_{n-1,j}(\f{y})}\right)\nonumber\\
&\stackrel{\eqref{eq:prop1}}{=}\left(\prod_{j=0}^{n-1} \left(g^{x_j}\right)^{\EF_{0,j}(\f{y})},\ldots,\prod_{j=0}^{n-1} \left(g^{x_j}\right)^{\EF_{n-1,j}(\f{y})}\right)
\label{eq:genfexp-raw-form}.
\end{align}

Hence, as in Section \ref{sec:fusexp}, we define
\begin{align*}
(g^{\f{x}})^{\f{y}}=(g^{(x_0,\ldots,x_{n-1})})^{(y_0,\ldots,y_{n-1})}\stackrel{\eqref{eq:genfexp-convention}}{=}(g^{x_0},\ldots, g^{x_{n-1}})^{(y_0,\ldots,y_{n-1})}
\end{align*}
through \eqref{eq:genfexp-raw-form} as
\begin{align}\label{eq:genfexp-preform}
(g^{\f{x}})^{\f{y}}:=\left(\prod_{j=0}^{n-1} \left(g^{x_j}\right)^{\EF_{0,j}(\f{y})},\ldots,\prod_{j=0}^{n-1} \left(g^{x_j}\right)^{\EF_{n-1,j}(\f{y})}\right)=g^{\f{x}\f{y}}.
\end{align}
Since we can write any element in $\G_q$ as a power of the primitive element $g$ we can set
$\f{g}=(g_0,\ldots,g_{n-1}):=(g^{x_0},\ldots, g^{x_{n-1}})$.
Substituting the powers of $g$ in \eqref{eq:genfexp-preform} gives
\begin{equation}\label{eq:genfexp-substituted}
\f{g}^{\f{y}}=\left(\prod_{j=0}^{n-1} g_j^{\EF_{0,j}(\f{y})},\ldots,\prod_{j=0}^{n-1} g_j^{\EF_{n-1,j}(\f{y})}\right)
\end{equation}
for any $\f{g}\in \G_p$ and $\f{y}\in \F_p$, fulfilling \eqref{eq:prop1} by construction.

So far we used a basis that is an $n$-tuple of elements of $\G_q$.
However, we did not yet constrain the basis elements.
To ensure that property \eqref{eq:prop2} holds, we need basis elements
from $\G_p$, being the direct product $\G_q^n$. Thus, for $\f{g},\f{h}\in \G_q$, multiplication in $\G_p$ is again component-wise
\begin{align}
(g_0,\ldots,g_{n-1})(h_0,\ldots,h_{n-1})=(g_0h_0,\ldots,g_{n-1}h_{n-1}).\label{eq:directprod}
\end{align}

The generalized construction fulfills \eqref{eq:prop2} by the linearity assertion \eqref{eq:lem_tilde}, because for any $\f{g}\in \G_p$ and $\f{x},\f{y}\in \F_p$ we have
\begin{align*}
\f{g}^{\f{x}+\f{y}}&\stackrel{\eqref{eq:genfexp-substituted}}{=}\left(\prod_{j=0}^{n-1} g_j^{\EF_{0,j}(\f{x}+\f{y})},\ldots,\prod_{j=0}^{n-1} g_j^{\EF_{n-1,j}(\f{x}+\f{y})}\right)\\
&\stackrel{\eqref{eq:lem_tilde}}{=}\left(\prod_{j=0}^{n-1} g_j^{\EF_{0,j}(\f{x})+\EF_{0,j}(\f{y})},\ldots,\prod_{j=0}^{n-1} g_j^{\EF_{n-1,j}(\f{x})+\EF_{n-1,j}(\f{y})}\right)\\
&\stackrel{\eqref{eq:prop2}}{=} \left(\prod_{j=0}^{n-1} \left(g_j^{\EF_{0,j}(\f{x})}g_j^{\EF_{0,j}(\f{y})}\right),\ldots,\prod_{j=0}^{n-1} \left(g_j^{\EF_{n-1,j}(\f{x})}g_j^{\EF_{n-1,j}(\f{y})}\right)\right)\\
&\stackrel{\eqref{eq:directprod}}{=}\left(\prod_{j=0}^{n-1} g_j^{\EF_{0,j}(\f{x})},\ldots,\prod_{j=0}^{n-1} g_j^{\EF_{n-1,j}(\f{x})}\right)\left(\prod_{j=0}^{n-1} g_j^{\EF_{0,j}(\f{y})},\ldots,\prod_{j=0}^{n-1} g_j^{\EF_{n-1,j}(\f{y})}\right)\\
&\stackrel{\eqref{eq:genfexp-substituted}}{=}\f{g}^{\f{x}}\f{g}^{\f{y}}.
\end{align*}

In the following examples are given for $n=1,2,3$. For simplicity we use a matrix representation for all equations with respect to \eqref{eq:prodcoeff} in the following manner for computing $\f{z}'=\f{x}\f{y}$, where $\f{x},\f{y}\in \F_p$:
\begin{align}
\f{z}'^T = \Lambda \f{x}^T\;\textrm{, where}\;\Lambda = \left(\EF_{i,j}(\f{y})\right)_{i,j=1}^n \in \F_q^{n\times n}\label{eq:matrix}
\end{align}
Accordingly, the following examples focus on the particular contents of $\Lambda$.

\begin{exl}
Let $n=1$, i.e. $\F_{q}=\Z/q\Z$. Then with respect to Equation \eqref{eq:matrix} we have $\Lambda = \left(y_0\right)$ which together with $g_0=g^{x_0}$ and Equation \eqref{eq:genfexp-substituted}, one gets
\begin{align*}
\f{g}^{\f{y}}\stackrel{}{=}g_0^{\EF_{00}(\f{y})}=g_0^{y_0}.
\end{align*}
\end{exl}
Notice that ordinary exponentiation is hence a special case of fusion exponentiation.
\begin{exl}
Let $n=2$, i.e. $\F_{q^2}=\Z_q[X]/(X^2+1)$. Then w.r.t.~Equation \eqref{eq:matrix} we have
\begin{align*}
\Lambda=
\left(\begin{array}{cc}
y_0 & -y_1\\
y_1 & y_0
\end{array}\right)
\end{align*}
which together with $g_j=g^{x_j}$, for $i,j=0,1$, and Equation \eqref{eq:genfexp-substituted} gives
\begin{align*}
\f{g}^{\f{y}}=\left(g_0^{y_0}g_1^{-y_1},g_0^{y_1}g_1^{y_0}\right).
\end{align*}
\end{exl}

\begin{exl}
Let $n=3$, i.e. $\F_{q^3}=\Z_q[X]/(X^3+X+1)$, for instance. Then w.r.t.~Equation \eqref{eq:matrix} we have
\begin{align*}
\Lambda=
\left(\begin{array}{ccc}
y_0 & -y_2 & -y_1\\
y_1 & y_0-y_2 & -y_2-y_1\\
y_2& y_1 & y_0-y_2
\end{array}\right)
\end{align*}
which together with $g_j=g^{x_j}$, for $i,j=0,1,2$, and Equation \eqref{eq:genfexp-substituted} gives
\begin{align*}
\f{g}^{\f{y}}&=\left(g_0^{y_0}g_1^{-y_2}g_2^{-y_1},g_0^{y_1}g_1^{y_0-y_2}g_2^{-y_2-y_1},g_0^{y_2}g_1^{y_1}g_2^{y_0-y_2}\right).
\end{align*}
\end{exl}

\paragraph{Fusion and Mixing:}
The concept of fusion has yet only been intuitively introduced by requiring a dependency of every component in the output on every component of the input. Similar concepts in cryptography exist, as for example the \emph{avalanche effect} calls for a similar influence on input bits on every output bit for a reasonable block-cipher. Here, things are slightly more involved, but the matrix structure may provide an answer on how the dependency relations look like. For example, if $\Lambda$ is of diagonal shape, then this results in a mere component-wise exponential function (cf. Equation (\ref{eq:genfexp-convention})). Otherwise, if the matrix is reducible, then its rows and columns can be permuted to reach a block-form, so that no cross-influence among blocks exist (a diagonal matrix is a trivial example). In the fusion exponentiation setting, this amounts to a failure of the desired mixing properties, as the set of input variables can be partitioned into at least two disjoint sets, with mutual influence present only within subsets, but not across all variables. Though a rigorous proof is yet not available, the matrix $\Lambda$ appears to never have zero entries and is as such always irreducible. It would follow that the desired dependencies exist among all variables, with no variable enjoying exceptionally stronger influence than any other.

\subsection{Resulting Definition of Generalized Fusion Exponential Function}

Since property \eqref{eq:prop3} is redundant (cf.~Remark \ref{rem:redundant}), we can state
\begin{defn}
Let $\F_p$ be a field with $p=q^n$, for some integer $n\geq 1$, and $\G_p$ be the $n$-fold direct product $\G_q^n$,
where $\G_q$ is a group of prime order $q$. The \emph{Generalized Fusion Exponential Function} is defined as
\begin{align}
\f{g}^{\f{x}}:=\left(\prod_{j=0}^{n-1} g_j^{\EF_{0,j}(\f{x})},\ldots,\prod_{j=0}^{n-1} g_j^{\EF_{n-1,j}(\f{x})}\right)\label{eq:def_gen_fusexp}
\end{align}\label{defn:genfusexp}
for $\f{g}\in \G_p$, $\f{x}\in \F_p$ and $\EF_{i,j}:\F_p\rightarrow \Z_q$, as defined in Section \ref{sec:vec_exp}.
\end{defn}

\begin{rem}
Notice that for $\f{g}=(g,1,\ldots,1)\in \G_p$ and $\f{1}=(1,0,\ldots,0)\in \F_p$ (i.e.~the 1-element in $\F_p$) we have
\begin{equation}
\f{g}=(g,1,\ldots,1)=(g^1,g^0,\ldots,g^0)\stackrel{\eqref{eq:genfexp-convention}}{=}g^{(1,0,\ldots,0)}=g^{\f{1}}.\label{eq:trivial_fexp}
\end{equation}
\end{rem}

\subsection{Primitive Elements}

In $\G_q$, the discrete logarithm of $y\in\G_q$ to the base $g\in \G_q$, $g\neq 1$, is
well defined because $q$ is prime. Since $|\Z_q|=|\G_q|$, exponentiation is bijective for exponents taken from $\Z_q$.
This property is important to keep computing discrete logarithms hard: any element $g\in \G_q$, distinct from $1$, is
a generator of $\G_q$. An analogous, and for many cryptosystems mandatory, result is that generalized fusion-exponentiation
is also bijective, thus, that any $\f{g}\in \G_p\setminus\{1\}$ can be used to generate the $n$-fold direct product $\G_p=\G_q^n$. However, by using the Generalized Fusion Exponential Function. In fact, this is true:

\begin{thm}
The Generalized Fusion Exponential Function is bijective.\label{thm:fusexpinj}
\begin{proof}
Since $|\G_p|=|\F_p|$, it suffices to show that the Generalized Fusion Exponential Function is injective.
Assume, that $\f{g}^{\f{b}}=\f{g}^{\f{c}}$ for some $\f{b},\f{c}\in \F_p$ and $\f{g}\in \G_p\setminus\{1\}$. Through \eqref{eq:genfexp-convention}, $\f{g}$ can
be written as $g^{\f{x}}$, for some $g\in \G_q\setminus\{1\}$, and some vector $\f{x}\in \F_p$, which, applied to $\f{g}^{\f{b}}=\f{g}^{\f{c}}$, gives $(g^{\f{x}})^{\f{b}}=(g^{\f{x}})^{\f{c}}$. Through property \eqref{eq:prop1} and the commutativity of multiplication in $\F_p$ we can write $(g^{\f{b}})^{\f{x}}=(g^{\f{c}})^{\f{x}}$, which holds if and only if $g^{\f{b}}=g^{\f{c}}$. By the injectivity of exponentiation in $\G_q$ this implies $b_i=c_i$ for all components $i=0,1,\ldots,n-1$ and hence $\f{b}=\f{c}$.
\end{proof}
\end{thm}
A consequence of this theorem is that given $\f{y}\in \G_p$ and $\f{g}\in \G_p\setminus\{1\}$ exactly one
$\f{x}\in \F_p$ exists, such that $\f{y}=\f{g}^{\f{x}}$. This justifies the following definition as sound:
\begin{defn}
Let $\f{g}\in \G_p\setminus\{1\}$. The \emph{Generalized Fusion Discrete Logarithm} is defined as follows:
\begin{align}
\textrm{fdlog}_{\f{g}}:\G_p\rightarrow \F_p,\quad \textrm{fdlog}_{\f{g}}(\f{y})=\f{x},\quad \textrm{s.t.}\;\f{y}=\f{g}^{\f{x}}
\end{align}
\end{defn}
\section{Fusion Discrete Logarithm Problems}\label{sec:fdlps}

In this section the Fusion Discrete Logarithm Problems are defined. Furthermore, some relations among these problems and the standard setting are shown.

\begin{defn}
Let $\F_p$, $\G_p$ and $n$ be as used in Definition \ref{defn:genfusexp} and assume that they are publicly known. Furthermore, let $\f{g}\in \G_p\setminus\{1\}$.\label{def:fdlp_family}
\begin{enumerate}
\item Let $\f{y}=\f{g}^{\f{x}}$, where $\f{x}\in\F_p$. The \emph{$n$-Fusion Discrete Logarithm Problem} ($n$-FDLP)
is the following: given $\f{y}$ and $\f{g}$, find
$\f{x}$.
\item Let $\f{y}_1=\f{g}^{\f{x}_1}$, $\f{y}_2=\f{g}^{\f{x}_2}$, where $\f{x}_1,\f{x}_2\in \F_p$. The \emph{$n$-Fusion Diffie-Hellman Problem} ($n$-FDHP) is the following: given $\f{y}_1$, $\f{y}_2$ and $\f{g}$, find $\f{g}^{\f{x}_1\f{x}_2}$.
\item Let $\f{y}_1=\f{g}^{\f{x}_1}$, $\f{y}_2=\f{g}^{\f{x}_2}$, $\f{y}_3=\f{g}^{\f{x}_3}$, where $\f{x}_1,\f{x}_2,\f{x}_3\in\F_p$. The \emph{$n$-Fusion Decision
Diffie-Hellman Problem} ($n$-FDDP) is the following: given
$\f{y}_1$, $\f{y}_2$, $\f{y}_3$ and $\f{g}$, decide if $\f{x}_3= \f{x}_1 \f{x}_2$.
\end{enumerate}
\end{defn}

\begin{figure}[h!]
\centering
\psfrag{A}[c][c][0.9][0]{\bfseries DLP}%
\psfrag{B}[c][c][0.9][0]{\bfseries DHP}%
\psfrag{C}[c][c][0.9][0]{\bfseries DDP}%
\psfrag{D}[c][c][0.9][0]{\bfseries $n$-FDLP}%
\psfrag{E}[c][c][0.9][0]{\bfseries $n$-FDHP}%
\psfrag{F}[c][c][0.9][0]{\bfseries $n$-FDDP}%
\psfrag{t}[c][c][0.9][0]{trivial}%
\psfrag{m}[c][c][0.9][0]{Thm.\ref{lem:DLP2DHP}, \cite{mw99}}%
\psfrag{u}[c][c][0.9][0]{unknown}%
\psfrag{x}[c][c][0.9][0]{Thm.\ref{lem:FDLP2DLP}}%
\psfrag{y}[c][c][0.9][0]{Thm.\ref{lem:FDHP2DHP}}%
\psfrag{z}[c][c][0.9][0]{Thm.\ref{lem:FDDP2DDP}}%
\psfrag{a}[c][c][0.9][0]{Cor.\ref{lem:FDLP2FDHP}}%
\psfrag{n}[c][c][0.9][0]{could yield benefits (cf.~Section \ref{sec:openprob})}%
\includegraphics{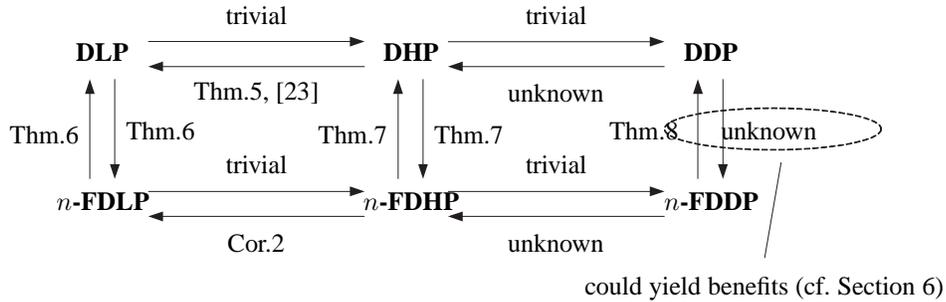}\caption{Relations among (Fusion) Discrete Logarithm
Problems, $n>2$.}\label{fig:fdlp_reduction}
\end{figure}

For reductions we use the following notation from complexity theory. Let $A$ and $B$ be two computational problems. We say that $A$ poly-time reduces to $B$ if an algorithm can be given which, using an oracle for $B$ as a subroutine, can solve $A$ with poly-time additional costs. This is denoted as A $\leq_P$ B. For the cases that $A \leq_P B$ and $B \leq_P A$ hold, we write $A \equiv_P B$ and say that ''$A$ and $B$ are computationally equivalent''.

Figure \ref{fig:fdlp_reduction} illustrates some relations among the Discrete Logarithm Problems in the standard and in the fusion-setting. Solving the $n$-FDLP leads to trivial solutions to the $n$-FDHP and the $n$-FDDP. Solving the $n$-FDHP leads to a trivial solution to the $n$-FDDP. Hence, we have an analogous result as in the standard setting.

As mentioned in the introduction, DLP $\equiv_P$ DHP has been shown to hold for groups whose order is free
of multiple large prime factors \cite{mw00}. This is the case for $\G_q$, since $q$ is prime. Having the trivial reduction DHP $\leq_P$ DLP we state the following theorem as a reference to the results in \cite{mw00}:

\begin{thm}
\emph{DLP} $\equiv_P$ \emph{DHP}\label{lem:DLP2DHP}
\end{thm}

For the reductions in the fusion-setting we start with the relation between the $n$-FDLP and the DLP.

\begin{thm}\label{thm:nfdlp-dlp-equivalence}
For all $n\in\N$ we have $n$-\emph{FDLP} $\equiv_P$ \emph{DLP}\label{lem:FDLP2DLP}
\begin{proof} 
For showing $n$-FDLP $\leq_P$ DLP, let $\f{y}=\f{g}^{\f{x}}$, where $\f{g}\in \G_p\setminus\{1\}$ and $\f{x}\in \F_p$. We wish to find $\f{x}$, given only $\f{y}$, $\f{g}$ and an oracle solving the DLP in polynomial time. Let $\f{g}=(g_0,\ldots,g_{n-1})=(g^{w_0},\ldots,g^{w_{n-1}})$ and $\f{y}=(y_0,\ldots,y_{n-1})=(g^{z_0},\ldots,g^{z_{n-1}})$ for some $g\in\G_q\setminus\{1\}$. We obtain $w_i=\textrm{dlog}_g(g_i)$ and $z_i=\textrm{dlog}_g(y_i)$, for $i=0,1,\ldots,n-1$, by using the oracle. Hence, we have
\begin{align*}
\f{g}^{\f{x}}\stackrel{\eqref{eq:genfexp-convention}}{=}\left(g^{\f{w}}\right)^{\f{x}}\stackrel{\eqref{eq:genfexp-preform}}{=}g^{\f{w}\f{x}}=g^{\f{z}}
\end{align*}
and thus $\f{z}=\f{w}\f{x}\in \F_p$. Since $\f{w}$, $\f{z}$
and $\F_p$ are known, one obtains $\f{x}=\f{z}\f{w}^{-1}$. Notice that $\f{w}\neq 0$ since $\f{g}\neq 1$.

For establishing DLP $\leq_P$ $n$-FDLP let $y=g^x$, where $g\in\G_q\setminus\{1\}$ and $x\in\mathbb{Z}_q$. We wish to find $x$, given only $y$, $g$ and an oracle solving the $n$-FDLP in polynomial time. Let $\textsf{y}:=(y,\ldots,y)\in \G_p$ and $\f{g}:=(g,1,\ldots,1)\in \G_p$. By \eqref{eq:trivial_fexp}, $\f{g}$ can be written
as $g^{\f{1}}$, where $\f{1}$ is the 1-element in $\F_p$. Furthermore, through \eqref{eq:genfexp-convention} $\f{y}$ can
be written as $g^{\f{x}}$, for $\f{x}:=(x,\ldots,x)\in \F_p$. Hence we have
\begin{align*}
\f{y}=g^{\f{x}}=g^{\f{1}\cdot\f{x}}\stackrel{\eqref{eq:genfexp-preform}}{=}\left(g^{\f{1}}\right)^{\f{x}}\stackrel{\eqref{eq:trivial_fexp}}{=}\f{g}^\f{x}
\end{align*}
and thus $\f{x}=\textrm{fdlog}_{\f{g}}(\f{y})$ can be obtained by the given oracle, revealing $x$.
Assuming the oracles are efficient, the above reductions are efficient too.
\end{proof}
\end{thm}
An immediate corollary is the following:
\begin{cor} For all $n,m\in\N\setminus\{0\}$, we have $n$-FDLP $\equiv_P$ $m$-FDLP.
\end{cor}
From a security-point-of-view, this means that the fusion setting is an asset in providing algebraic properties, but will not give increased security by hardening any underlying computational problem. We come back to this later, when we discuss possible applications.

Based on Theorems \ref{lem:DLP2DHP} and \ref{lem:FDLP2DLP} we can state
\begin{thm}
For all $n\in\N$ we have, $n$-\emph{FDHP} $\equiv_P$ \emph{DHP}\label{lem:FDHP2DHP}
\begin{proof}
Due to Theorem \ref{lem:DLP2DHP} we have DLP $\equiv_P$ DHP and together with Theorem \ref{lem:FDLP2DLP} we thus have
\begin{equation*}
\textrm{DHP}\equiv_P\textrm{DLP}\equiv_Pn\textrm{-FDLP}\geq_Pn\textrm{-FDHP}.
\end{equation*}
For the reverse direction DHP $\leq_P$ $n$-FDHP let $y_i=g^{x_i}$, where $x_i\in\mathbb{Z}_q$, for $i=1,2$. Querying an oracle for the $n$-FDHP with the inputs
$\textsf{y}_1=(y_1,1,\ldots,1)$, $\textsf{y}_2=(y_2,1,\ldots,1)$ and $\textsf{g}=(g,1,\ldots,1)$ results in
$\textsf{y}_3=(g^{x_1x_2},1,\ldots,1)$, since $(y_1,1,\ldots,1)=(g,1,\ldots,1)^{(x_1,0,\ldots,0)}$,
$(y_2,1,\ldots,1)=(g,1,\ldots,1)^{(x_2,0,\ldots,0)}$ and $(x_1,0,\ldots,0) (x_2,0,\ldots,0)=(x_1x_2,0,\ldots,0)$. Thus, $y_3$ is stored in the first component of $\textsf{y}_{3}$.
\end{proof}
\end{thm}

The result $n$-FDHP $\equiv_P$ DHP together with DLP $\equiv_P$ DHP and DLP $\equiv_P$ $n$-FDLP gives the same relation between the $n$-FDLP and the $n$-FDHP as in the standard setting, summarized in

\begin{cor}
For all $n\in\N$, we have $n$\emph{-FDLP} $\equiv_P$ $n$\emph{-FDHP}\label{lem:FDLP2FDHP}
\end{cor}

Regarding the DDP and the $n$-FDDP the situation is less clear. The following theorem shows the trivial reduction from the DDP to the $n$-FDDP.

\begin{thm}
For all $n\in\N$, we have \emph{DDP} $\leq_P$ $n$\emph{-FDDP}\label{lem:FDDP2DDP}
\begin{proof}
Let $y_i=g^{x_i}$, where $x_i\in\mathbb{Z}_q$, for $i=1,2,3$. Notice that $(x_1,0,\ldots,0)
(x_2,0,\ldots,0)=(x_1x_2,0,\ldots,0)$. Hence, querying an oracle for the
$n$-FDDP with the inputs $\textsf{y}_1=(y_1,1,\ldots,1)$, $\textsf{y}_2=(y_2,1,\ldots,1)$, $\textsf{y}_3=(y_3,1,\ldots,1)$ and $\textsf{g}=(g,1,\ldots,1)$
results in 1, iff $(x_3,0,\ldots,0)=(x_1,0,\ldots,0) (x_2,0,\ldots,0)$, and 0 otherwise.
\end{proof}
\end{thm}
The reverse direction is unknown and might yield some security benefits (cf.~Section \ref{sec:openprob}). Also it is unkown whether $n$-FDHP $\leq_P$ $n$-FDDP holds (as in the the standard setting).

\begin{rem}
Notice that the bit-security is always
associated to the same prime $q$, since the standard and the fusion-setting refer to the same security parameter $q$. Thus, the $n$-FDLP, regardless of how large $n$ is, can never be harder than the DLP. The attacks always work with running time in $O(\sqrt{q})$.
\end{rem}
\section{Possible Security Benefits}\label{sec:openprob}

One interesting open problem is to show $n$-FDDP $\leq_P$ DDP, for $n>1$ (of course $n=1$ is trivial since $1$-FDDP = DDP). Since we want to find a generic algorithm, we are only allowed to use the group operations as black-boxes and an oracle for solving the DDP in polynomial time. Such an orcale, however does not provide more than
\textsf{true}/\textsf{false}-decisions. All current
approaches to give an efficient reduction to the DDP end up in the
necessity to have an oracle for solving the DHP. Such an oracle,
however, is not available for this (direct) reduction from the $n$-FDDP to the DDP (i.e.~without solving the DHP or DLP). 

The above stated open problems yield an interesting conjecture: if
 the computational equivalence between the DDP and the $n$-FDDP cannot be
 shown for \emph{all} $n>1$, then the $n$-FDDP seems to be a stronger problem than the DDP (at least for one $n$). Thus, if the DDP is efficiently solved directly (i.e.~without solving the
 DLP or DHP), then related cryptosystems like ElGamal or Cramer-Shoup will become vulnerable. However, if our conjecture remains unrefuted, then such
cryptosystems will still
 remain secure within the (generalized) fusion-setting. 
\section{Applications}

It is obvious that the fusion-setting is less efficient than the standard setting. With $n$ the number of exponentiation in $\G_q$ increase with quadratic complexity. Asides from the possible security benefits as stated in section \ref{sec:openprob} the following applications might be of interest:

\paragraph{Verifiable Secret Sharing in $\F_{q^n}$:} Shamir's secret sharing scheme \cite{s79} is normally used for sharing secrets in $\Z_q$. It is secure against $t<n$ passive adversaries. If the holder of a share sends a corrupted value during the reconstruction phase the result is incorrect. To counter this problem mechanisms can be included to enable all participants to jointly identify active malicious parties. Such sharing schemes are called Verifiable Secret Sharing. Many of them make use property of \eqref{eq:prop2} of exponentiation in $\G_q$ such that the verification of shares can be done in hidden form. For the case that one wants to share a secret in $\F_{q^n}$ then all such Verifiable Secret Sharing schemes can be transferred to the fusion-setting, since Fusion Exponentiation provides the same property and security level.

\paragraph{Security Multi-Party Computation in $\F_{q^n}$ with Active Adversaries:} Secure multi-party computation over shared secrets in $\Z_q$ is well known \cite{gmw87,grr98,h01}. The protocols with security against passive adversaries (like it is the case for Shamir's secret sharing) are generic in the way that they can also be applied if secrets are shared in $\F_{q^n}$. For protocols being secure against active adversaries, verifiable secret sharings schemes as the ones mentioned above are often used. Using Fusion Exponentiation again yields the benefit that security multi-party computation over $\F_{q^n}$ with respect to active adversaries can be realized using the fusion-setting.

\paragraph{Threshold Cryptosystems in $\G_{q^n}$:} Clearly, DL-based cryptosystems can be realized in the fusion-setting. Due to the fact that verifiable secret sharing and secure multi-party computation can be used straightforwardly in $\F_{q^n}$, transforming DL-based threshold cryptosystem to the fusion-setting is easy.

\paragraph{Signature Schemes:} 
Apart from the well-known concept of signature, such as put forth in the first papers about public-key cryptography, a vast amount of more sophisticated concepts has evolved. As for instance, \emph{redactable signatures} \cite{Johnson2002} allow for exchanging certain parts of a document without invalidating a signature. Aggregate signatures \cite{Boneh2003} permit assembly of several signatures into a single one, multisignatures \cite{Itakura1983} are the several-person-pendant to a standard signature, and so on. As most of these are based on arithmetics that has been carried over to the fusion-setting, fusion exponentiation appears as a natural candidate for constructing signatures with modifiable components, or with several signatures being aggregated, yet still verifiable one by one.
\section{Future Work}

The full potential of fusion exponentiation is for sure not exhaustively described by this paper. Among the open problems (which may yield security benefits compared to the ordinary setting) is a formalization of the \emph{fusion properties} (i.e. dependencies of output variables on input variables), and their connection to the structure of the matrix $\Lambda$. This one may be the key for proving a property that is known as avalanche effect in different contexts. Even more interesting is the potential for constructing sophisticated signature schemes, that otherwise (until now) rely on more complicated algebraic structures like supersingular hyperelliptic curve groups and bilinear pairings. Finally, the concept opens is fascinating from a purely algebraic point of view too, since it appears to be the first generalization of the exponential function that carries over to vectors in the exponent in finite fields. 
\appendix
\section{Further Example-Instantiations of Fusion-Exponentiation}

\begin{exl}
Let $n=4$, i.e. $\F_{q^4}=\Z_q[X]/(X^4+X+1)$, for instance. Then w.r.t.~Equation \eqref{eq:matrix} we have
\begin{align*}
\Lambda=
\left(\begin{array}{cccc}
y_0 & -y_3 & -y_2 & -y_1 \\
y_1 & y_0-y_3 & -y_2-y_3 & -y_1-y_2\\
y_2 & y_1 & y_0-y_3 & -y_2-y_3 \\
y_3 & y_2 & y_1 & y_0-y_3
\end{array}\right)
\end{align*}
which together with $g_j=g^{x_j}$, for $i,j=0,\ldots,3$, and Equation \eqref{eq:genfexp-substituted} gives
\begin{align*}
\f{g}^{\f{y}}
=&\left(g_0^{y_0}g_1^{-y_3}g_2^{-y_2}g_3^{-y_1},g_0^{y_1}g_1^{y_0-y_3}g_2^{-y_2-y_3}g_3^{-y_1-y_2},g_0^{y_2}g_1^{y_1}g_2^{y_0-y_3}g_3^{-y_2-y_3},g_0^{y_3}g_1^{y_2}g_2^{y_1}g_3^{y_0-y_3}\right).
\end{align*}
\end{exl}

\begin{exl}
Let $n=5$, i.e. $\F_{q^5}=\Z_q[X]/(X^5+X^2+1)$, for instance. Then w.r.t.~Equation \eqref{eq:matrix} we have
\begin{align*}
\Lambda=
\left(\begin{array}{ccccc}
y_0 & -y_4 & -y_3 & -y_2 & -y_1+y_4 \\
y_1 & y_0 & -y_4 & -y_3 & -y_2 \\
y_2 & y_1-y_4 & y_0-y_3 & -y_2-y_4 & -y_1-y_3+y_4\\
y_3 & y_2 & y_1-y_4 & y_0-y_3 & -y_4-y_2\\
y_4 & y_3 & y_2 & y_1-y_4 & y_0-y_3
\end{array}\right)
\end{align*}
which together with $g_j=g^{x_j}$, for $i,j=0,\ldots,4$, and Equation \eqref{eq:genfexp-substituted} gives
\begin{align*}
\f{g}^{\f{y}}
=&\left(g_0^{y_0}g_1^{-y_4}g_2^{-y_3}g_3^{-y_2}g_4^{-y_1+y_4},g_0^{y_1}g_1^{y_0}g_2^{-y_4}g_3^{-y_3}g_4^{-y_3},\right.\\
&\qquad\left.g_0^{y_2}g_1^{y_1-y_4}g_2^{y_0-y_3}g_3^{-y_2-y_4}g_4^{-y_1-y_3+y_4},g_0^{y_3}g_1^{y_2}g_2^{y_1-y_4}g_3^{y_0-y_3}g_4^{-y_4-y_2}\right).
\end{align*}
\end{exl}

\bibliographystyle{plain}
\bibliography{FDLP}

\end{document}